\documentclass[conference]{IEEEtran}

\usepackage{times}
\usepackage{latexsym}
\usepackage{epsfig}
\usepackage{subcaption}

\usepackage[T1]{fontenc}

\usepackage[utf8]{inputenc}

\usepackage{microtype}

\usepackage{inconsolata}

\usepackage{graphicx}

\usepackage{amsmath,amssymb,amsfonts}
\usepackage{color}
\usepackage{listings,mdframed}

\usepackage{booktabs}
\usepackage{pifont}
\usepackage{array}
\newcommand{\cmark}{\ding{51}}

\usepackage{xspace}
\usepackage{enumitem}
\newcommand{\name}{GenCC\xspace}
\newcommand{\evolve}{evolve\xspace}
 
\newcommand{\CoT}{Math-CoT\xspace}

\definecolor{mygreen}{rgb}{0,0.5,0}
\definecolor{myred}{rgb}{0.8,0,0}
\definecolor{myblue}{rgb}{0,0,1}
\newcommand{\liron}[1]{{\color{mygreen}{#1}}}
\newcommand{\neta}[1]{{\color{myblue}{#1}}}

\newcommand{\hide}[1]{}
\newcommand{\hidearxiv}[1]{}
\newcommand{\todo}[1]{{\color{red}{#1}}}

\title{Utility Function is All You Need: LLM-based Congestion Control}

\author{\IEEEauthorblockN{Neta Rozen-Schiff}
\IEEEauthorblockA{\textit{TU Berlin}}
\IEEEauthorblockA{\textit{Germany}}
\and
\IEEEauthorblockN{Liron Schiff}
\IEEEauthorblockA{\textit{Akamai Technologies}}
\IEEEauthorblockA{\textit{USA}}
\and
\IEEEauthorblockN{Stefan Schmid}
\IEEEauthorblockA{\textit{TU Berlin, Weizenbaum Institute \& Fraunhofer SIT}}
\IEEEauthorblockA{\textit{Germany}}
}

\begin{document}
\maketitle
\begin{abstract}
Congestion is a critical and challenging problem in communication networks. Congestion control protocols allow network applications to tune their sending rate in a way that optimizes their performance and the network utilization. In the common  distributed setting, the applications can’t collaborate with each other directly but instead obtain similar estimations about the state of the network using latency and loss measurements. These measurements can be fed into analytical functions, referred to by utility functions, whose gradients help each and all distributed senders to converge to a desired state. 

The above process becomes extremely complicated when each application has different optimization goals and requirements. Crafting these utilization functions has been a research subject for over a decade, with small incremental changes requiring rigorous mathematical analysis as well as real-world experiments.

In this work, we present \name, a framework leveraging the code generation capabilities of large language models (LLMs) coupled with realistic network testbed, to design congestion control utility functions.
Using \name, we analyze the impact of different guidance strategies on
the performance of the generated protocols, considering application-specific requirements and network capacity. 

Our results show that LLMs, guided by either a generative code evolution strategy or  mathematical chain-of-thought (CoT), can obtain close to optimal results, improving state-of-the-art congestion control protocols by $37\%$-$142\%$, depending on the scenario.

\end{abstract}

\section{Introduction}


Today, Internet access is possible in almost any corner of the world, but a reliable and high-bandwidth connection is not available to everyone even in developed countries—a reality that many readers may be experiencing firsthand when participating in remote conference calls.
While some users benefit from high-capacity fiber links, others contend with the limitations of broadband or cellular asymmetric access.  The rapid expansion of the Internet, its massive user base, and the increasing reliance on broadband connectivity have therefore made congestion one of the most critical challenges in networking \cite{congestion_challenge, utility_per_network_cond}. 

Congestion arises when the traffic demand on a network link exceeds its available capacity. Efficiently reconciling such disparities by optimizing bandwidth allocation across shared bottlenecks has been a central challenge for congestion control (CC) research for decades \cite{BBR,LEDBAT_rfc,PCC,cubic,vivace}.

 
Many classic CC solutions struggle to meet the growing demands of modern network applications \cite{streaming_review}. 
Moreover, these applications increasingly combine multiple functionalities—including control, telemetry, video, and audio, 
each imposing distinct and sometimes conflicting performance requirements on the network. For example, in interactive video services, the control channel typically requires low bandwidth and low latency, whereas $4$K video streams demand high bandwidth while tolerating moderate latency \cite{PRISM}.

To address these challenges, recent CC protocols such as Proteus \cite{proteus} and Hercules \cite{Hercules} 
use an online learning approach which models the network as a repeated multi-user game. In this game, the users repeatedly decide on their sending rates which are expected to converge to an optimal equilibrium state.
While these protocols can achieve high performance, they rely on carefully engineered utility functions and nontrivial parameter tuning, which limits adaptability in dynamic and heterogeneous network environments \cite{DL_TCP1, DL_TCP2}. Moreover, these designs assume that a single utility function remains appropriate across varying network conditions—an assumption that does not hold in practice \cite{utility_per_network_cond}.

Large language models (LLMs) have demonstrated strong capabilities in code generation \cite{LLM_for_code2,LLM_for_code1}, opening new opportunities to revisit the design space of CC and to explore more flexible, automated approaches for utility-function generation. 



We design \name, a framework leveraging these opportunities by implementing different strategies of LLM-piplines coupled with network testbed which can experiment and measure the quality of the generated functions.
Our pipelines use different guidance strategies ranging from simple zero-shot prompting to mathematical chain-of-thought (CoT)\cite{chain_of_thought}. 
Moreover, we evaluate the impact of an evolutionary process that leverages the network testbed to provide feedback to the LLM, guiding it to generate improved functions.

In summary, our key contributions are fourfold:
\begin{itemize}
    \item We leverage the mathematical framework and architecture of utility-based congestion control protocols, together with the mathematical reasoning and coding capabilities of LLMs, to design enhanced and improved CC protocols.
    
    \item{We explore different LLM guidance strategies and compare 
    their impact on the performance of the generated protocols.}
    
    \item We implement \name, an infrastructure that uses LLMs and a realistic testbed to generate CC protocols for specific scenarios involving connections with heterogeneous requirements. \name can also be used as a benchmark for different LLM models.
    \item We evaluate the \name-generated CC protocols on live broadband and inter-cloud traffic, achieving up to a $2.4×$ throughput improvement over the state-of-the-art.
\end{itemize}

As a contribution to the research community, to ensure reproducibility and to facilitate follow-up work, we will release all our code as open source. 
\section{Background and Goals}\label{sec:background}


Recent research has shifted toward improving Quality of Service (QoS) for heterogeneous traffic, based on the  observation that ``the era of friendly TCPs has been broken down'' \cite{FB_TCP}, as emerging applications increasingly prioritize performance objectives beyond fairness. 

To this end, two state-of-the-art control protocols were developed Proteus \cite{proteus} and Hercules \cite{Hercules}, both are based on an \emph{online learning} approach which minimizes \emph{a priori} assumptions about the network and has been shown to robustly achieve high performance across a wide range of challenging environments.

\begin{figure*}
  \centering
  \includegraphics[width=0.7\linewidth]{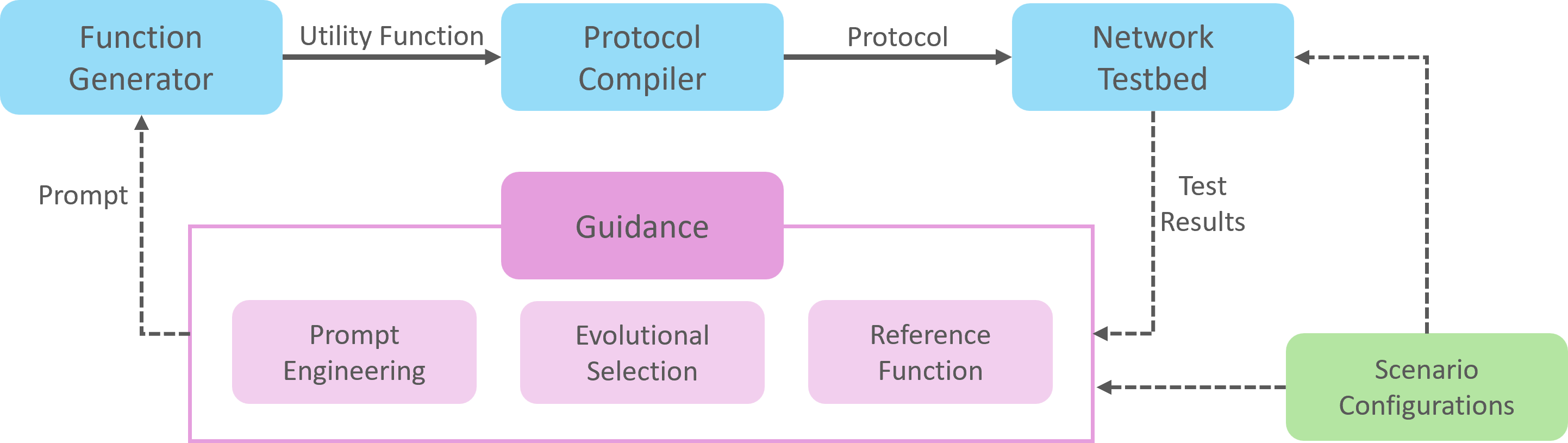}
\caption{\name's cyclic pipeline} 
\label{fig:Genn_cc_process}
\end{figure*}

\vspace{0.05in} \noindent{\bf Utility function is all we need.}
In the online learning approach, time is partitioned into discrete intervals. During each interval $t$, a connection transmits at a constant sending rate $r_t$ (its current strategy) and receives feedback from the receiver in the form of selective acknowledgments (SACKs). These acknowledgments are used to compute performance statistics such as loss rate and latency (RTT). The resulting statistics, together with $r_t$ are provided as inputs to a utility function in order to calculate $r_{t+1}$ (the next interval's strategy) based on the change in utility, i.e., its gradient \cite{vivace,PCC}.
\hide{
This repeated calculation, performed in parallel by all senders, can be seen as multivariate gradient descent and can converge to equilibrium state 
depending on the utility function used.
Therefore, the definition of the utility function determines the scenarios in which equilibrium will be achieved, the speed of convergence, and the conditions under which guarantees can be provided. It also dictates whether network-wide properties (e.g., fairness and utilization) are ensured and whether per-connection requirements are taken into account.
}

This iterative procedure, executed concurrently by all senders, can be viewed as a form of multivariate gradient descent. 
Consequently, 
the definition of the utility function determines the scenarios in which equilibrium is achieved, the rate of convergence, and the guaranties that can be established. These guaranties may include network-wide properties (e.g., fairness and utilization) \cite{vivace} as well as satisfaction of per-connection requirements \cite{Hercules}.
However, these guaranties are achieved through parameter tuning to specific scenarios and traffic patterns.

\vspace{0.05in}
\noindent{\bf Can a single utility function fit all network scenarios?}
Most CC protocols use a fixed utility function \cite{cubic,FB_TCP,vivace,MPCC,proteus}. Recent work shows that this is often suboptimal, and that adapting utility-function \emph{parameters} to the \emph{network context} can significantly improve performance. Network context includes application objectives (e.g., latency or video quality), temporal load variations, and deployment scenarios such as $5G$ versus LTE \cite{utility_per_network_cond}.


Motivated by the importance of network context—and by evidence that even modest changes to utility-function parameters can yield significant performance gains—\name~goes beyond parameter tuning and focuses on automatically \emph{generating} the most suitable utility function for a given network scenario using LLMs. This naturally raises a key question: what form of \emph{guidance} should be provided to the LLM to enable effective and stable utility-function generation? We address this question in the following sections.

\section{\name Overview}\label{sec:overview}

\name is a framework for generating customized CC protocols using LLMs.
\name supports different guidance strategies, representing common and advanced prompt engineering techniques as well as an evolutionary code optimization process. 
For all guidance strategies, \name uses a similar cyclic pipeline, described in Fig.~\ref{fig:Genn_cc_process}, which involves generating utility functions and integrating them into a full
 CC protocol. 
To this end,  \name adapts and extends implementations of existing utility based CC protocols, 
leveraging the fact that such protocols use the utility function as a black box. 


\subsection{Guidance}\label{sec:guidance}

While LLMs are very capable in reasoning and code generation tasks, using them in state-of-the-art applications and research often requires more context and guidance in order to obtain quality results \cite{utility_per_network_cond}. Moreover, generating multiple instances of code that can be automatically integrated and compiled as part of existing code requires accurate instructions and constrains.

\begin{table*}[t]
\vspace{-0.35cm}
\centering
\small
\setlength{\tabcolsep}{6pt}
\begin{tabular}{
>{\raggedright\arraybackslash}p{1.4cm}
>{\centering\arraybackslash}p{1cm}
>{\centering\arraybackslash}p{1.1cm}
>{\centering\arraybackslash}p{0.9cm}
>{\raggedright\arraybackslash}p{9.2cm}}
\toprule
\textbf{Strategy} & \textbf{Example} &  \textbf{Feedback} & \textbf{CoT} & \textbf{Description} \\
\midrule

Zero-shot
&  &  &  
& Simple prompt requesting a function for congestion control. \\

One-shot
& \cmark &    &
& Similar to zero-shot, but with Hercules utility function as an example. \\

%
Math-CoT
&  &  & \cmark
& Extending zero-shot with analytical properties and steps. \\

%
Evolve
& \cmark  & \cmark &
& Select the best (so far) and try to improve it given its weakest connection.\\

\bottomrule
\end{tabular}
\caption{\name's guidance strategies.}
\vspace{-0.35cm}
\label{table:guidance_settings}
\end{table*}

In this work we explore four guidance strategies: zero-shot, one-shot, Math-CoT and evolutionary (`Evolve'). These strategies are summarized in Table~\ref{table:guidance_settings} and implemented by the following sub-modules:

\vspace{0.05in}\noindent{\bf Prompt Engineering} is used in all strategies, we ask the LLM to generate a utility function for CC, provide the explicit C++ function header and, depending on strategy, we may add extra context in the form of an example function that needs to be improved. While in most strategies we use a straightforward prompt, in the mathematical chain-of-thought (Math-CoT) strategy we elaborate on the mathematical properties of the desired function (e.g., variables' meaning) that can help to analytically verify it. \hidearxiv{The exact prompts are provided in detail in Appendix~\ref{sec:appendix}.}

\vspace{0.05in}\noindent{\bf Evolutionary selection.} 
Across all strategies, we use the least satisfied connection as the primary performance metric. As detailed in Section~\ref{sec:evaluation}, performance is averaged over multiple tests per utility function, and the best function is selected per experiment, where an experiment consists of a sequence of consecutive generations.
In the \evolve~ strategy, however, the performance observed during the experiment is used to adapt the prompt by providing the best function identified so far together with its performance (i.e., the least satisfied connection requirement). Unlike the other strategies, which rely on a fixed prompt and LLM non-determinism to obtain improved functions, \evolve~ applies an evolutionary selection mechanism to progressively refine the generated functions.


\vspace{0.05in}\noindent{\bf Reference function} is used both as an example for the one-shot strategy and for initialization (first generation) of the \evolve strategy. 
We select  the state-of-the-art CC protocol Hercules \cite{Hercules} as a reference due to its demonstrated ability to support a large number of heterogeneous connection requirements.

\hide{
For each connection $i$, the input includes the parameters $(a_i, b_i)$, the minimum and maximum requirements respectively. 
Then, the sending rate $x_i$ is normalized accordingly by: 
$\overline{x_i} = \frac{x_{i}-a_{i}}{b_{i}-a_{i}}$. Based on the normalized sending rate,
Hercules' utility function is defined as follows \cite{Hercules}:
\begin{align}\label{eq:Hercules_utility}
    U(x_{i}) =& (x_{i})^t -  x_{i} \cdot H(\overline{x_i}) \cdot [\beta \cdot L_{i} \nonumber \\
    &
     + \gamma \cdot \max\{0,\frac{d(RTT_{i})}{dt}\} 
     + \varphi \cdot \sigma(RTT_i)], \nonumber
\end{align}
where, $L_i$ is the $i^{th}$ connection loss rate, $RTT_i$ represents its RTT gradient and  $H(\overline{x_i})$ is a scaling factor of this connection, which defined as follows:
\begin{align} \label{eq:H(x)}
    H(\overline{x_i}) &= \frac{\arctan(D \cdot  (\overline{x}_{i} - \frac{1}{2}))}{\pi}+\frac{1}{2} \in (0,1), \nonumber
\end{align}
}

The input for any utility function includes the the minimum and maximum requirements $(a, b)$, the current sending rate $x$, and the measured loss rate $L$, RTT gradient $\frac{d(RTT)}{dt}$ and variance $\sigma(RTT)$. Hercules' utility function, $U$, uses the normalized sending rate  
$\overline{x} = \frac{x-a}{b-a}$ and is defined as follows:
\begin{align}
    U(x) =& x^t -  x \cdot H(\overline{x}) \cdot [\beta \cdot L \nonumber \\
    &
     + \gamma \cdot \max\{0,\frac{d(RTT)}{dt}\} 
     + \varphi \cdot \sigma(RTT)], \nonumber
\end{align}
where, $H(\overline{x})$ is a scaling factor defined by:
\begin{align} \label{eq:H(x)}
    H(\overline{x}) &= \frac{\arctan(D \cdot  (\overline{x} - \frac{1}{2}))}{\pi}+\frac{1}{2} \in (0,1), \nonumber
\end{align}

\subsection{Function Generator}
Function generation is performed by invoking an LLM based on the prompt provided by the Guidance module. In this work, we use GPT-$5$ for a balance between performance and cost; however, \name~ supports any locally or remotely hosted model accessible via a compatible REST API.

In our experiments we found that explicit request from the LLM to receive only the function code was followed correctly in almost all responses, obviating the need for further output processing\footnote{Note that in critical systems it is advised to perform code sanitation and even sandboxing before compiling and running generated code in the wild.}.


\subsection{Protocol Compiler}
\name builds upon and extends the experimental implementation of Proteus \cite{Proteus_code}, which includes a simple client and server that execute the protocol and report per-run throughput and latency.

Leveraging the modular nature of utility-based protocols, \name encapsulates the utility function as a standalone module, enabling rapid replacement without modifying the rest of the code base. This design also facilitates direct integration of the LLM-generated output with minimal adjustment; compilation failures occurred only in rare cases (discussed in Section~\ref{sec:performance}). Since the utility function operates solely at the sender, only the client binary must be updated.


\hide{
\vspace{0.05in}\noindent{\bf Code compilation.}
\name's extension enables each utility function to be implemented as a standalone module with a well-defined interface that receives a predefined set of input parameters as described above. At this stage, the utility function is compiled and prepared for testing.
}

\subsection{Network Testbed}
The network testbed executes the compiled protocol client (with a fixed server) for each configured connection, as specified in the requirements configuration file. For each setup, it runs a predefined number of tests (default: $10$), each lasting a fixed duration (default: $30$ seconds), after which the results are collected and aggregated.

The testbed also accepts a scenario-specific network script defining bandwidth, latency, jitter, burst size, and packet loss constraints, which are enforced using the Linux \emph{tc} tool.

Experiments can be conducted either from a local machine, reflecting real Internet streaming conditions, or from a cloud instance with high baseline bandwidth, where constraints are applied to induce congestion.

\section{Evaluation}\label{sec:evaluation}

\begin{figure*}
 \centering
\begin{subfigure}{0.27\linewidth}
  \centering
  \includegraphics[width=\linewidth]{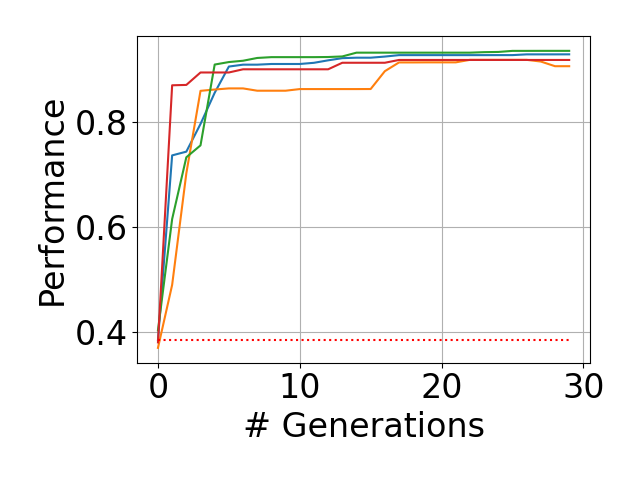}
  \caption{Broadband}
  \label{fig:cable_generation}
\end{subfigure}%
\begin{subfigure}{0.27\linewidth}
  \centering
  \includegraphics[width=\linewidth]{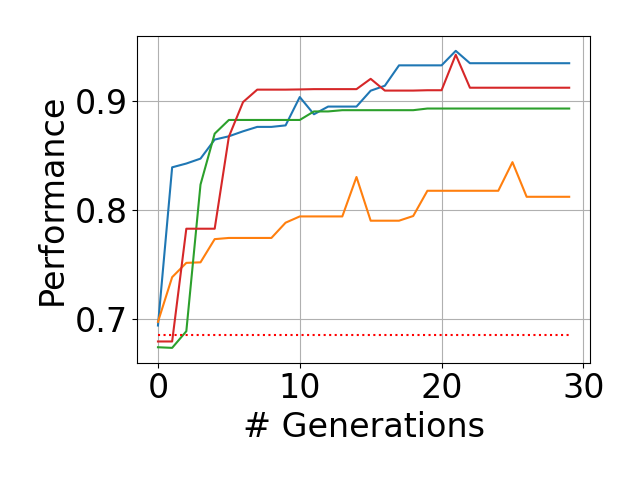}
  \caption{Cellular}
  \label{fig:5G_generation}
\end{subfigure}%
\begin{subfigure}{0.27\linewidth}
  \centering
  \includegraphics[width=\linewidth]{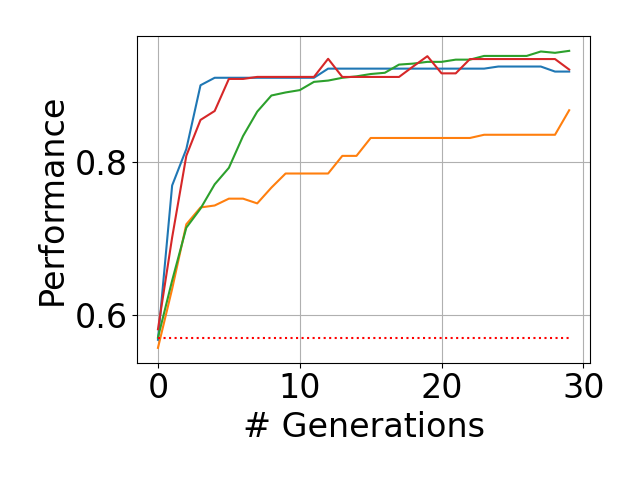}
  \caption{Satellite}
  \label{fig:Leo_generation}
\end{subfigure}%
\begin{subfigure}{0.25\linewidth}
  \centering\includegraphics[width=\linewidth]
 {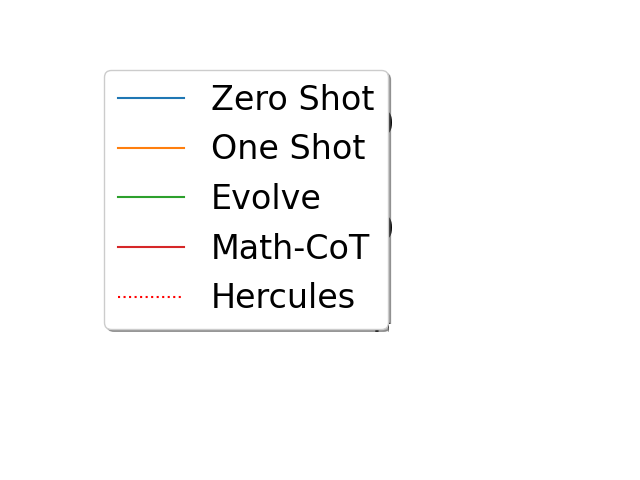}
  \end{subfigure}%
\caption{Performance improvement as with function generation in broadband, cellular and satellite scenarios}
\label{fig:AWS_loss}
\vspace{-0.5cm}
\end{figure*}

\subsection{Methodology}\label{sec:Methodology}
We use \name to evaluate the impact of the guidance strategy on the performance of the generated protocols and to compare them to the state-of-the-art congestion control protocol Hercules \cite{Hercules}. To the best of our knowledge, these protocols are the only ones to explicitly address heterogeneous connection requirements. Each protocol is evaluated with constrained connections between two cloud machines (constrained-cloud scenarios) and unconstrained connections between a local machine and a cloud machine (in-the-wild scenario).


\vspace{0.05in}\noindent{\bf Constrained-cloud Scenarios}
consist of two Akamai Cloud dedicated CPU instances\cite{linode}, a sender in Chicago and a receiver in Atlanta, where the outgoing traffic of the sender is shaped to obtain three representative network scenarios:

\begin{enumerate}[noitemsep,topsep=2pt,leftmargin=*]
\item \textbf{Satellite:} $10$~Mbps bottleneck capacity and $60$~ms RTT, reflecting the upper bound of common LEO satellite uplink connections in the U.S.~\cite{Leo_Ookla,INGR_2023}.
\item \textbf{Cellular:} $20$  Mbps bottleneck and $40$~ms RTT, aligning with the upper bound of experienced $5$G cellular uplink performance in the U.S.~\cite{5G_Ookla} and the FCC cellular benchmark~\cite{5G_FCC}.
\item \textbf{Broadband:} $50$~Mbps capacity and $20$~ms RTT, representative of typical residential broadband cable uplink connections in the U.S., and close to average wired broadband performance (DSL and fiber)~\cite{Cable_FCC}.
\end{enumerate}

\vspace{0.05in}\noindent{\bf In-the-wild scenario}
consists of a local machine located at Boston, representing home user or office workstation, connected through WiFi router and broadband WAN to the Internet. From that local machine we create connections to a cloud instance in Atlanta. These connections experience similar network conditions as the constrained high bandwidth scenario but not as fixed and stable.


\begin{figure}
  \centering
  \includegraphics[width=0.65\linewidth]{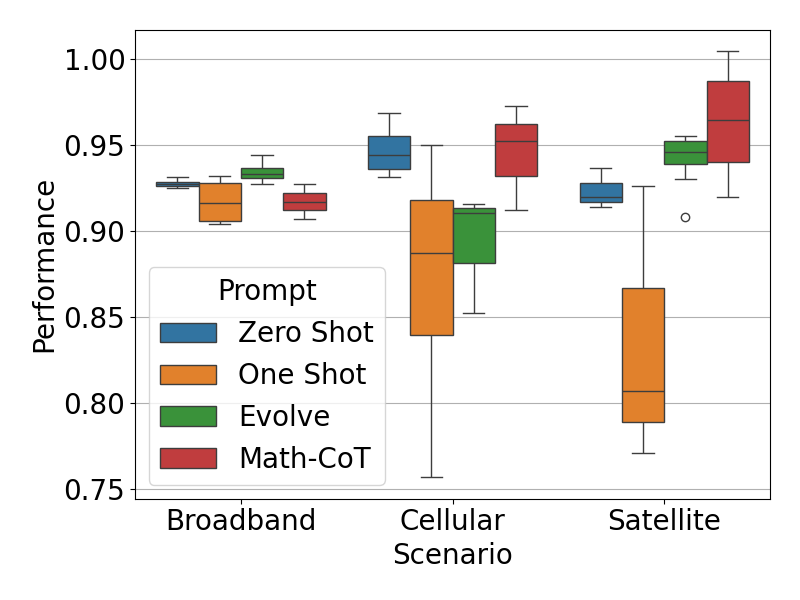}
  \vspace{-0.5cm}
\caption{Performance in constrained cloud scenarios} 
\label{fig:cloud_constrained}
\vspace{-0.5cm}
\end{figure}

\vspace{0.05in}\noindent{\bf Requirements}
of modern applications often exceed the available network resources \cite{Congestion_Leo,5G_throughput_latency,congestion_challenge} and comprise multiple functional connections, at least one of which imposes bandwidth demands that surpass the capacity of the underlying network \cite{Hercules}. 

To further stress-test our evaluation, we evaluate each scenario using a combination of five  heterogeneous connections whose minimum requirements differ by an order of magnitude. This wide range of requirements spans applications from low-rate IoT traffic to highly video workloads, which challenge the tested networks.
The maximum requirement for each connection is set to $1.5\times$ its minimum requirement, providing a $50\%$ margin.

\subsection{Performance}\label{sec:performance}

\vspace{0.05in}\noindent{\bf Definition.}
Following previous work \cite{Hercules}, we consider the \emph{''satisfaction ratio``} provided by a CC protocol for a connection $i$ in a given scenario, as the ratio between the average sending rate (in that scenario) and the minimum requirement ($a_i$) of that connection. 
Moreover, we consider the worst case (i.e., guaranteed) satisfaction ratio of a protocol as the minimum satisfaction ratio across all connections.
Therefore, we define the performance of a protocol (or utility function) in a scenario as the ratio between the average worst case satisfaction across all repetitions (default $10$) divided by the optimal solution in that scenario.

\begin{figure}
  \centering
  \includegraphics[width=0.65\linewidth]{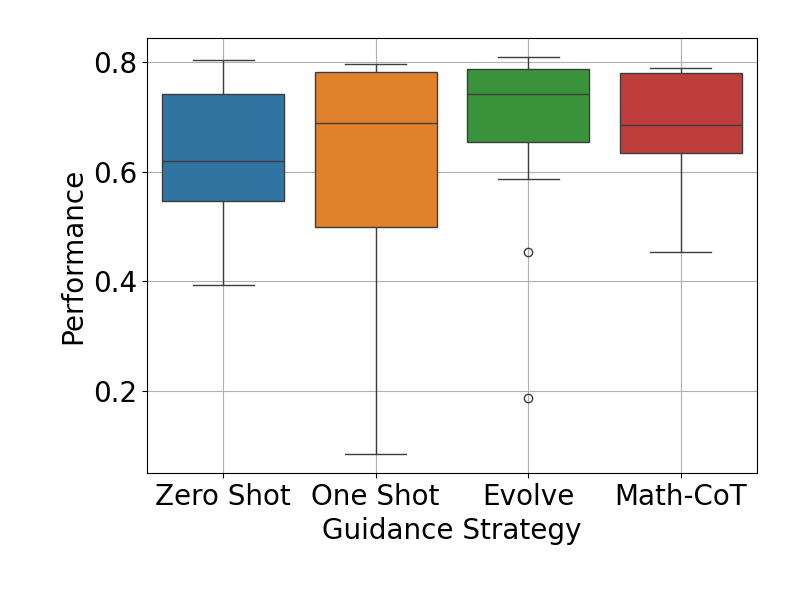}
  \vspace{-0.5cm}
\caption{Performance for the in the wild scenario} 
\label{fig:home_cable_optimal}
\vspace{-0.5cm}
\end{figure}



\vspace{0.05in}\noindent{\bf Comparison to the state-of-the-art.}
Fig.~\ref{fig:AWS_loss} presents the performance of the best utility function identified up to each generation, evaluated across the different scenarios in the cloud-constrained experiments. The results indicate that zero-shot, \evolve~and \CoT strategies outperform one-shot strategy by $5\%$–$10\%$. Specifically, \evolve~and \CoT strategies  significantly surpass the state-of-the-art congestion control protocol, Hercules, achieving performance gains of $142\%$ in the broadband scenario, $37\%$ in the cellular scenario, and $67\%$ in the satellite scenario.

In addition, the leading strategies reach their best-performing function within $10$ trials per scenario, whereas the one-shot approach requires about $20$ generations to achieve peak performance. Given the relatively small number of generations needed, we henceforth focus on the best function obtained in each experiment, treating it as an upper bound on each guidance strategy’s capability.

\vspace{0.05in}\noindent{\bf Providing more information can harm.}
Fig.~\ref{fig:cloud_constrained} and Fig.~\ref{fig:home_cable_optimal} present the best-function average performance and diversity for each LLM guidance strategy across $10$ experiments, for the cloud-constrained and in-the-wild broadband settings, respectively.

\begin{figure}
 \vspace{-0.5cm}
  \centering
  \includegraphics[width=0.65\linewidth]{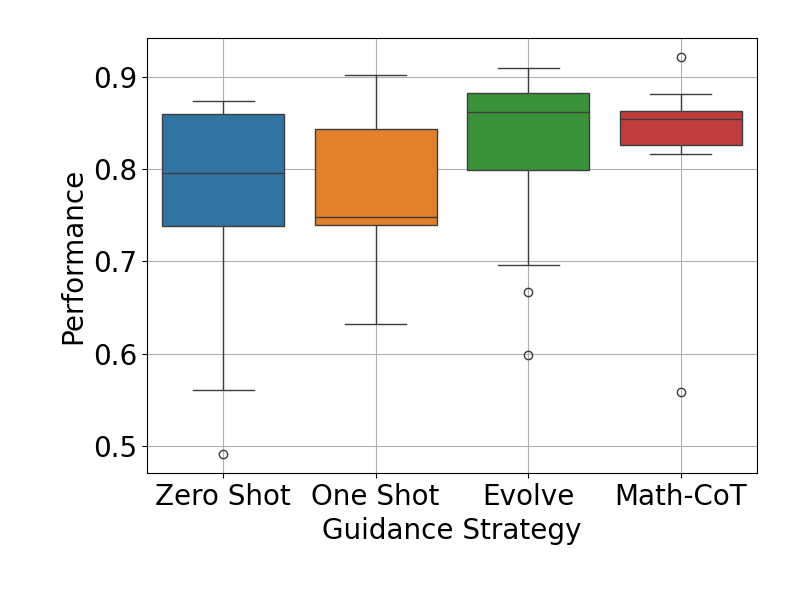}
  \vspace{-0.5cm}
\caption{Cross scenario performance} 
\label{fig:other_scenarios}
\vspace{-0.5cm}
\end{figure}
The results show that the one-shot strategy performs worst in the satellite and cellular scenarios and exhibits a large standard deviation in the broadband setting. These findings indicate that providing additional information in the prompt does not necessarily improve performance; in particular, repeatedly supplying the same reference function leads to underperforming and unstable outcomes, even compared to the zero-shot approach.

As expected, performance in the cloud-constrained setting exceeds that of the in-the-wild broadband scenario. Notably, however, the relative ranking of the strategies remains consistent across both settings. 
\emph{Overall, the best-performing strategies are \evolve and \CoT.} 

\vspace{0.05in}\noindent{\bf Validity of the function code.}
Under the \evolve\ and \CoT\ strategies, all generated utility functions were valid. In contrast, the one-shot approach produced approximately $1$–$2\%$ invalid functions (depending on the scenario), while the zero-shot approach resulted in about $1\%$ invalid functions in the broadband scenario only. The corresponding plot is omitted due to space constraints.

We also evaluated a lightweight open-source model (GPT-OSS–$20$B). However, across all guidance strategies (\evolve, zero-shot, \CoT, and one-shot), approximately $80\%$ of the generated outputs failed to compile.

\vspace{0.05in}\noindent{\bf Fairness.} We evaluate utility functions' fairness in two in-the-wild broadband experiments with identical requirements. Each experiment includes two connections with identical requirements vary between $20$ Mbps minimum requirement to $75$ Mbps maximum requirement. 
All generated protocols achieved fair-share between the two connections. 
Plots are omitted due to space constraints.

\vspace{0.05in}\noindent{\bf Cross scenarios performance.}
We further evaluated the constrained-cloud scenarios to assess the generality of the best functions. Specifically, the top-performing function from each scenario was tested on the other two. The results show that \evolve\ and \CoT\ achieve nearly $90\%$ of the optimal performance on average, while zero-shot reaches $80\%$ and one-shot $75\%$ (see Fig.~\ref{fig:other_scenarios}).

\section{Related Work}

\vspace{0.05in}\noindent{\bf Congestion control} has long been formulated as an optimization problem \cite{TCP_improvment,CC_old_survey}. Traditional protocols assume similar requirement for among all the connection sending traffic and enforce fairness using loss and RTT based utility functions, such as CUBIC \cite{cubic}, Vegas \cite{Wvegas}, COPA \cite{COPA}, Vivace \cite{vivace}, and BBR \cite{BBR}.

More recent approaches incorporate traffic heterogeneity by extending utility functions with additional parameters, such as RTT deviation  \cite{proteus,Hercules}. However, using a single utility function across all network conditions remains suboptimal; adapting utility parameters to network context and application requirements can significantly improve performance \cite{utility_per_network_cond}.

This observation let us to try generate utility function with network context, by using LLMs.

\vspace{0.05in}\noindent{\bf LLM for Congestion Control.}
Recent work has explored using LLMs to improve CC, including selecting CC algorithms and tuning TCP parameters based on network conditions \cite{LLM_TCP}, as well as applying LLM-based code generation to enhance BBR \cite{Microsoft_BBR}. However, these approaches do not account for network context or heterogeneous traffic requirements, and therefore can lead to bandwidth over allocated for low demand flows and insufficient resources for high demand ones as discussed in \cite{PRISM}.

\section{Conclusion}

In this paper, we present \name, a framework that enhances congestion control protocols through LLM-guided utility-function generation. \name integrates network context and heterogeneous application requirements and supports multiple guidance strategies, including zero-shot, CoT, and an evolutionary approach that feeds back the best-performing function at each iteration. The latter two strategies improve state-of-the-art performance by $37\%$–$142\%$.

We further show that selecting effective LLM guidance is nontrivial. In particular, repeatedly providing the same reference utility function (one-shot) can degrade performance and increase instability compared to zero-shot guidance.

\section{Acknowledgment} This research was supported by the German Research Foundation (DFG), Schwerpunktprogramm: Resilienz in Vernetzten Welten (SPP $2378$), project ReNO, $2023-2026$.

\bibliographystyle{IEEEtran}
\bibliography{LLM_for_CC.bib}

\end{document}